\documentclass[prl,aps,showpacs,lengthcheck]{revtex4}
\usepackage{amsmath}
\usepackage{color}
\usepackage{amsfonts}
\usepackage{graphicx}

\newcommand{\bq}{\begin{equation}} 
\newcommand{\eq}{\end{equation}}
\newcommand{\ba}{\begin{eqnarray}}
\newcommand{\ea}{\end{eqnarray}}

\begin{document}

\title{Mechanically driven growth of quasi-two dimensional microbial colonies}

\author{F.~D.~C. Farrell$^1$, O. Hallatschek$^{2, 3}$, D. Marenduzzo$^1$, B. Waclaw$^1$} 
\affiliation{$^1$SUPA, School of Physics and
  Astronomy, University of Edinburgh, Mayfield Road, Edinburgh EH9
  3JZ, UK \\
  $^2$MPI for Dynamics and Self-Organization, Bunsenstrasse 10, D-37073 G{\"o}ttingen, Germany \\
  $^3$Department of Physics, University of California, Berkeley, CA 94720, USA}

\begin{abstract}
We study colonies of non-motile, rod-shaped bacteria growing on solid substrates.  In our model, bacteria interact purely mechanically, by pushing each other away as they grow, and consume a diffusing nutrient. We show that mechanical interactions control the velocity and shape of the advancing front, which leads to features that cannot be captured by established Fisher-Kolmogorov models. In particular, we find that the velocity depends on the elastic modulus of bacteria or their stickiness to the surface. Interestingly, we predict that the radius of an incompressible, strictly two-dimensional colony cannot grow linearly in time. Importantly, mechanical interactions can also account for the nonequilibrium transition between circular and branching colonies, often observed in the lab.
\end{abstract}

\maketitle

%introduction

Active matter, which constantly takes energy from its environment in order to do work~\cite{ramaswamy}, has recently attracted much interest.
Particular examples are collections of cells such as tissues and suspensions of swimming bacteria~\cite{goldstein-swimmers,instab-swimmers,epithelial}, and microbial colonies, in which activity is caused by growth, death and migration of cells. The combination of these three factors has been shown to lead to a variety of interesting and universal patterns \cite{ben-jacob_generic_1994, pnasdav, branching, bonachela_universality_2011}. For example, bacteria such as {\em B. subtilis} or {\em E. coli} grown on Petri dishes form patterns ranging from circular, through Eden-like \cite{eden}, to diffusion-limited aggregation-like patterns \cite{fujikawa_fractal_1989}. Such patterns have been traditionally modelled using a system of diffusive Fisher-Kolmogorov equations \cite{ben-jacob_cooperative_2000, murray} which combine migration (diffusion of bacteria), bacterial growth, and nutrient diffusion. 
This approach, however, does not accurately represent the growth on surfaces on the microscopic level, where expansion is caused by cells pushing each other out of the way as they grow, rather than by migration.

\begin{figure}
\begin{minipage}{7cm}
\includegraphics[width=3.3cm]{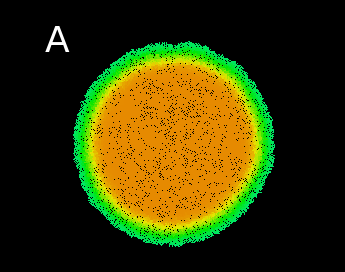}
\includegraphics[width=3.3cm]{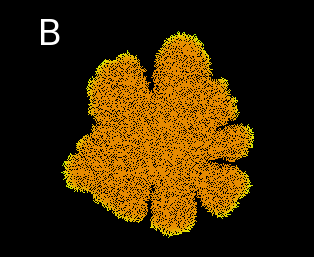}
\includegraphics[width=3.3cm]{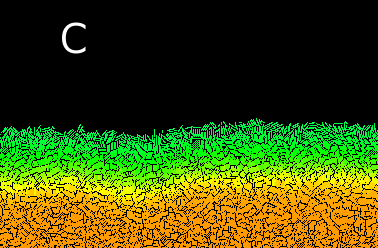}
\includegraphics[width=3.3cm]{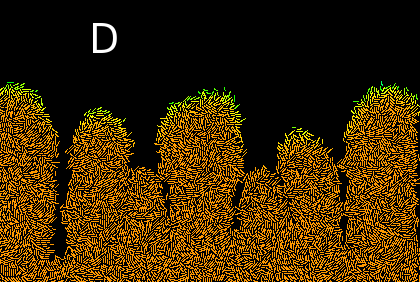}
\end{minipage}
\begin{minipage}{1cm}
\includegraphics[width=0.7cm]{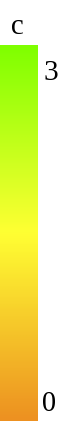}
\end{minipage}
\includegraphics[width=0.6\columnwidth]{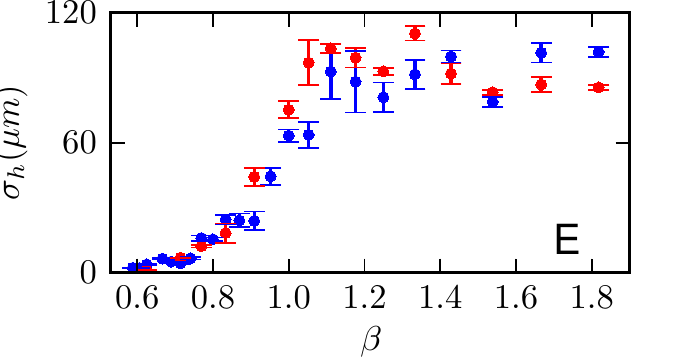}
\caption{Top panel: snapshots from the simulation of $N \sim 10^5$ cells, for low (A) and high (B) values of the branching parameter $\beta$. 
Colours correspond to the local nutrient concentration, see the the colormap on the right. Only a thin layer of cells (green) grows appreciably. Middle panel: growth in a narrow, long strip, for low (C) and high (D) $\beta$; only the growing layer is shown. The frame is co-moving with the front. Bottom panel: roughness of the front as a function of $\beta$, for cells with maximal aspect ratio 4:1 (blue, {\em E. coli}-like) and 2:1 (red, {\em S. cerevisiae}-like). See Supplemetal Material for typical parameter values used. }\label{shot}
\end{figure}   

In this paper, we study the role of mechanical interactions between cells in the growth of  dense colonies on solid substrates. Inspired by recent experiments in microfluidic devices \cite{tsimring}, we study a simple problem of quasi-two dimensional growth of a colony of {\it non-motile} single-celled organisms which consume nutrient in order to grow and divide. We argue -- supported by computer simulations and analytical calculations -- that mechanical interactions between bacterial cells can account for  the emergence of a nonequilibrium transition between quasi-circular and branched colonies as a function of the ratio between the nutrient consumption rate and the growth rate. An effectively density-dependent consumption rate, postulated in the Fisher equation framework~\cite{murray}, arises naturally in our model due to compressibility of cells or their escape into the third dimension (forming multiple layers). The strength of mechanical interactions determines the speed with which the colony expands in space, with diffusion of the nutrient playing a secondary role. We also show that the leading edge of the front is very sharp, and the bacterial density is discontinuous at the front, in contrast to a smooth, exponential profile predicted by models based on coupled Fisher equations~\cite{murray,branching}. 
Our results are relevant to the growth of biofilms \cite{costerton_bacterial_1999, xavier_social_2009, hense_does_2007}, which are ubiquitous in nature and are involved in a variety of medical and technological problems. As mechanical interactions may alter the colony morphology, and the fixation probability of (potentially harmful) mutants \cite{hallatschek_life_2010, kuhr_range_2011}, understanding their role is of paramount importance.

%Simulations
We simulate bacteria using two-dimensional Newtonian dynamics. Cells are modelled as growing spherocylinders of constant diameter $d=1\mu$m and variable length that split in half to yield two cells when they reach some critical size (which varies slightly from cell to cell). The colony grows on a two-dimensional flat surface with nutrient concentration $c(x,y)$. The nutrient diffuses with diffusion constant $D$. Nutrients are consumed at a rate $k f(c)$ per unit biomass density, where $f(c)$ is a monotonously increasing dimensionless function of $c$. In most simulations, we use a Monod function $c/(c_{\rm half}+c)$ with half-saturation constant $c_{\rm half}$. Cells grow (by elongation) at a rate $\phi f(c)$. The typical values of all parameters are detailed in the Supplemental Material.

The cells interact mechanically in a similar way to that of Ref.~\cite{tsimring,boyer_buckling_2011}. The force between overlapping bacteria is assumed to be given by the Hertzian theory of elastic contact~\cite{landauelasticity}: $F=E d^{1/2}h^{3/2}$ where $h$ is the overlap and $E$ parametrizes the strength of the interaction and is proportional (modulo a dimensionless prefactor) to the elastic modulus of the cells. We also assume that the dynamics is overdamped so that the velocity of a cell is proportional to the force exerted on it (see Supplemental Material).

We start our simulations from either a single initial cell or a line of cells, and follow the shape of the colony after many rounds of cell replication, leading to a circular colony or a horizontal advancing front, respectively. Figure 1 shows that the morphology of a large colony of bacteria can be either smooth or branched, depending on the parameters of the model.  

By performing simulations for different parameter sets we have found that the fate (smooth/branched) of the colony is determined by a dimensionless ``branching parameter'' $\beta=(k \rho_0)/(\phi c_0)$, where $c_0$ is the initial nutrient concentration, $\rho_0$ the densely-packed cell density, and the other parameters have been defined previously. 

For small values of $\beta$, the front of the colony remains smooth throughout the simulation (Fig. 1A,C), whereas for large values branches develop (Fig. 1B,D). Note that, as in real colonies~\cite{hallatschek_life_2010}, the nutrient becomes depleted within the colony so that only cells in a thin layer at the front are growing. To pinpoint the location of the transition more accurately, we compute the roughness of the front (Fig. 1E), defined as the mean square deviation of points on the front from its average position, as in Ref.~\cite{bonachela_universality_2011}. The roughness measured at steady state increases by over an order of magnitude as $\beta$ passes through $1.0$. The transition between planar and branching front is largely independent of the aspect ratio of the cells (Fig. 1E). 

This transition between branched and smooth colony fronts is well known in real colonies~\cite{shapiro}, and has been the subject of many theoretical studies~\cite{murray,branching}, which usually attribute it to the interplay between diffusion (migration) of bacteria and diffusion of the nutrient. In our model, however, the transition is driven by the uptake of nutrient by the cells and their growth by mechanical pushing, and is unaffected by the diffusion rate of the nutrient. 

To gain a better understanding of the physics of this transition, we approximate the growing colony as an incompressible cellular ``fluid'' \footnote{A similar analysis has been performed in Ref.~\cite{klapper_finger_2002} in the context of biofilm growth.}. Mass conservation in such a fluid is described by the equation $\nabla\cdot\mathbf{v}=\phi f(c(\mathbf{x}))$, where $\mathbf{v}$ the fluid velocity, $\phi$ is the growth rate, and $f(c)$ is the dimensionless nutrient uptake function. This is coupled to an equation describing the diffusion and depletion of the nutrient. Let us begin with a one dimensional case of a colony advancing from the left and characterized by a single number $x_0(t)$ which is the position of the front:
\ba
 \partial_t c(x,t) &=& D\partial_x^2 c(x,t) -k \rho_0 f(c(x,t)) \Theta(x_0-x), \label{diff} \\
	v(x_0) &=& \frac{dx_0}{dt}=\phi \int_{-\infty}^{x_0(t)}f(c(x,t))dx.	\label{front}
\ea
Here $D$ is the nutrient diffusion constant, $k$ the rate of uptake of nutrient by cells, $\rho_0$ the cell density (constant everywhere due to incompressibility), and $\Theta$ is the Heaviside step function. Because cells do not migrate and they are tightly packed, the density is either $\rho_0$ or zero, and hence equation~(\ref{front}) can be derived from the continuity equation and the incompressibility condition, assuming that $\rho(x,t)=\rho_0 \Theta(x_0(t)-x)$. We also impose boundary conditions that $c(-\infty)=0$ and $c(\infty)=c_0$. 

We first determine whether Eqs. (\ref{diff},\ref{front}) admit a travelling-wave solution $c(x,t)=\hat{c}(x-vt)\equiv \hat{c}(z)$ in the limit $t\to\infty$, where the velocity $v$ of the front is constant. The resulting equations for $\hat{c}(z)$ and the front velocity $v$ are   
\ba
 -v\hat{c}'(z) &=& D\hat{c}''(z)-k\rho_0 f(\hat{c})\Theta(-z) \label{sofc}, \\
	v &=& \phi \int_{-\infty}^{0}f(\hat{c}(z))dz \label{x0}.
\ea
For $z>0$, it is easily seen that the solution to Eq.~(\ref{sofc}) is given by $\hat{c}(z)=c_0+Ae^{-vz/D}$ (as $c(\infty)=c_0$). For $z<0$, we can rearrange the equation to yield $f(\hat{c}(z))=\frac{1}{k\rho_0}\left(D\hat{c}''(z)+v\hat{c}'(z))\right)$, which, upon insertion into Eq.~(\ref{x0}) gives
\bq
	v =	\frac{\phi}{k\rho_0} \left(D\hat{c}'(0)+v\hat{c}(0)\right) = \frac{\phi c_0}{k\rho_0}v,
\eq
where we have integrated by parts, and used the fact that $\hat{c}$ vanishes at $-\infty$, and that $\hat{c}$ and $\hat{c}'$ must be continuous at $z=0$. Therefore, a solution for $v$ exists only if $\phi c_0 = k \rho_0$ (or $\beta=1$) exactly: we have found that in the incompressible limit the front cannot advance at a constant speed! This is in contrast to the Fisher framework, where travelling waves exist for a range of parameters.  Numerical solutions of Eqs.~(\ref{diff}, \ref{front}) fully confirm our prediction, see Supplemental Material.

\begin{figure}
\includegraphics[width=0.32\columnwidth]{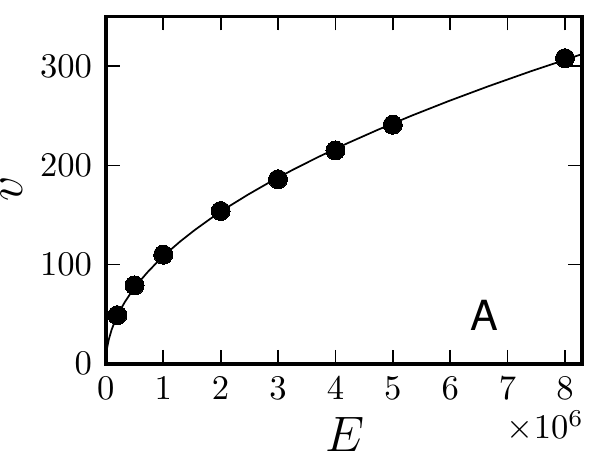}
\includegraphics[width=0.32\columnwidth]{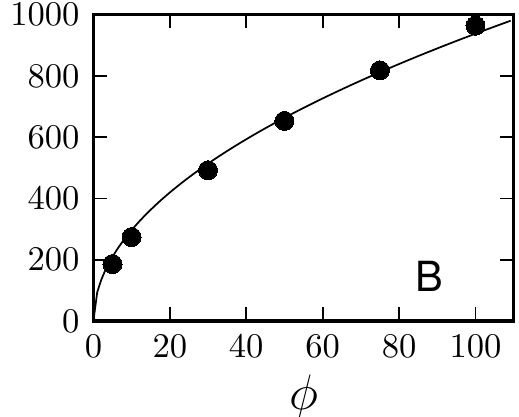}
\includegraphics[width=0.32\columnwidth]{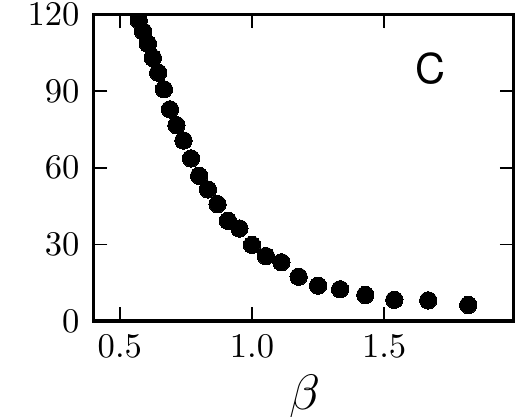}
\caption{Steady state speed of colony growth, $v$, as a function of various parameters, for 2D simulations in the quasi-1D geometry. A and B have fits to a square root function. In B, $\phi$ is varied while holding $\beta$ constant (by varying $k$). C shows the dependence on $\beta$, with a change in behaviour around $\beta=1$.}
\end{figure}

The hint from this simplified 1D model is therefore that $\beta=1$ is a critical value that separates different regimes of colony growth. For $\beta>1$, growth is limited by the nutrient diffusion rate, whereas for $\beta<1$ diffusion does not play any role. However, there are two problems here. First, the front has more freedom in 2D than in 1D - it can become branched and the profile does not have to be circularly symmetric. Since this change occurs around $\beta=1$, it is therefore appealing to conjecture that the morphological transition in Fig. 1 is linked to the switch in growth laws described above in the theory for an incompressible colony. 

%velocity in quasi-1D simulations and experiments -- effect of compressibility

Second, incompressible theory predicts that under no conditions can growth be linear, unless $\beta=1$. This is inconsistent with experimental results: the size of a colony of non-swimming bacteria growing on stiff agar gels {\em does} increase linearly with time~\cite{hallatschek_genetic_2007}. Moreover, our simulations also lead to a finite steady state speed. The speed found in simulations depends on the elasticity $E$, as can be seen in Fig. 2A, suggesting the compressibility of the cells is important. 

Generalizing the theory above to compressible cells, we now need equations for mass and momentum conservation, as well as the nutrient diffusion equation, still in the 1D geometry:
\ba
 \partial_tc &=& D\partial_x^2c-k \rho f(c) \label{cm1}, \\
 \partial_t\rho+\partial_x(\rho v) &=& \phi \rho f(c) \label{cm2}, \\
 \partial_x p &=& - \mu \rho v \label{cm3}.
\ea
The term $\mu \rho v$ describes the friction between the surface and the cells. The pressure $p(\rho)$ is determined by the force acting between the cells. We take $p(\rho(x))=E(1-\rho_0/\rho(x))^{3/2}$ to be consistent with our simulations, because the force that acts between two overlapping cells is then proportional to $E d^{1/2} h^{3/2}$, where $h=d(1-\rho_0/\rho(x))$ is the overlap.

Although Eqs.~(\ref{cm1}-\ref{cm3}) cannot be solved analytically, a numerical solution (see Supplemental Material) shows that a travelling wave now exists for $\beta<1$. The density profile close to the edge decays according to a power law towards the uncompressed cell density $\rho_0$. This power law decay and the finite density at the very edge are in striking contrast to Fisher-Kolomogorov waves, which exhibit exponential density profiles in the wave tip \cite{murray}. Many other properties of the solution to Eqs.~(\ref{cm1}-\ref{cm3}) can be deduced without solving the equations. First, a ``biomass conservation law'' from Eqs.~(\ref{cm1}) and (\ref{cm2}) states that one unit of nutrient biomass makes $\phi/k$ units of bacterial biomass, and hence the density $\rho(-\infty)$ deep in the colony must be $\phi c_0/k$. This explains why a travelling wave solution cannot exist in the incompressible case: unless the cell density $\rho_0$ equals {\it exactly} $\phi c_0/k$ it will not match the density of biomass produced by the nutrient. It also explains why there is a morphological transition to branched colonies at $\beta\simeq 1$: growth of a flat front is not possible for $\beta>1$ as it would need to have a density less than $\rho_0$.  Finally, it suggests that if bacteria are restricted to grow as a monolayer, then, when nutrient is abundant, they will grow exponentially until intermicrobial forces within the colony are so large that the bacteria in the middle are squashed to the appropriate density $\rho_0/ \beta$. 

The idea that the cell population has to be compressed to a normal strain of $\epsilon \equiv 1-\beta$ for the front to grow at a constant speed can be turned into a simple scaling argument. At steady state the pressure profile has to rise from $0$ at the edge of the population to a maximal value $p_*$ in the bulk within a boundary layer of characteristic size $\lambda$. The characteristic length $\lambda$ can be eliminated by estimating it to be the length by which the front moves in one generation $\lambda \approx v /(\phi f(c_0))$. The bulk value of the pressure $p_*(\epsilon)$ is just large enough that the density of the population is compressed down to the strain $\epsilon$. The elastic constitutive relation $p_*(\epsilon)$ of the microbial population fixes the corresponding pressure, with $p_*(\epsilon)=E \epsilon^{3/2}$ in our case of Hertzian contacts between cells. The pressure $p_*$ pushes the front population at the speed $v$ against the friction force $v \mu \rho_0 \lambda$, where $\mu \rho_0$ acts as a friction coefficient per unit length. Force balance thus yields
\begin{equation}
  \label{eq:lambda-v-compressible}
  v\approx \sqrt{\frac{\phi f(c_0) p_*(\epsilon)}{\mu \rho_0}}=\sqrt{\frac{E \phi f(c_0) }{\mu \rho_0}} g(\beta),
\end{equation}
where $g(\beta)=(1-\beta)^{3/4}$.

\begin{figure}
\includegraphics[width=0.49\columnwidth]{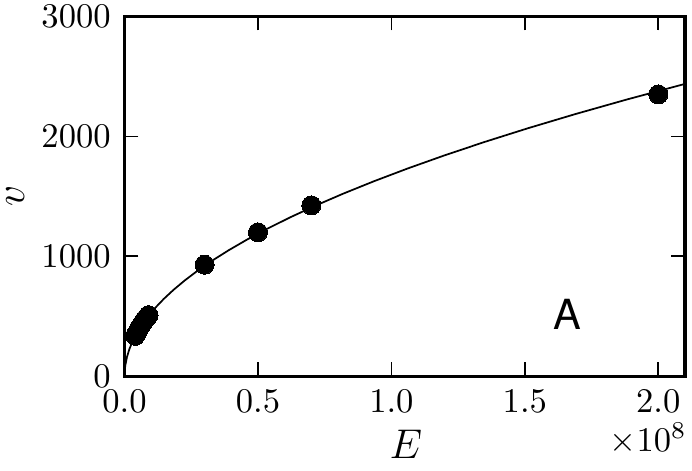}
\includegraphics[width=0.49\columnwidth]{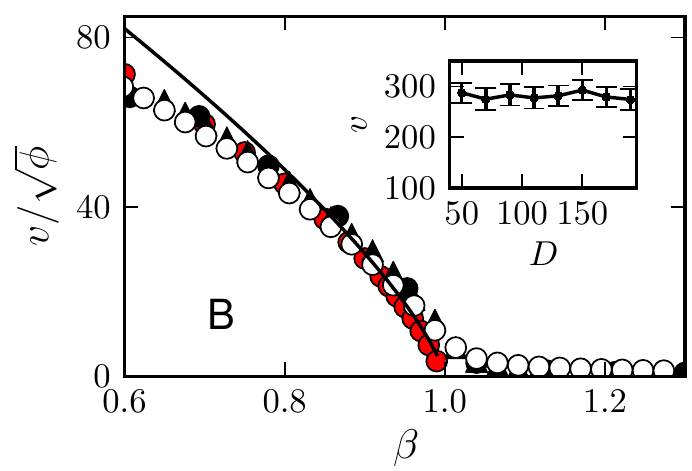}
\caption{\label{1d}Dependence of front speed on parameters in the fully 1D simulation. A: front speed as a function of repulsion strength $E$, with fit to $v=A\sqrt{E}$. B: Transition from moving to stopped front as a function of $\beta$, which occurs when $\beta=1$. $v/\sqrt{\phi}$ is plotted against $\beta$ (by varying $k$) for $\phi=$ 10 (open circles), 20 (triangles) and 30 (closed circles), showing that $v \sim \sqrt{\phi} g(\beta)$. Here $E=4\times10^6$, $D=100$. Solid line corresponds to theoretical $g(\beta)=(1-\beta)^{3/4}$, and red (grey) circles are the numerical solution of Eqs.~(\ref{cm1}-\ref{cm3}). Inset: $v$ as a function of $D$, showing no dependence.} 
\end{figure}

To test the above formula, we performed a fully one-dimensional version of our simulations described above, as this removed the effects of branching and was much more computationally efficient. The results are shown in Fig.~\ref{1d}. Figure~\ref{1d}A shows that the front speed grows as $\sqrt{E}$ as predicted by Eq.~(\ref{eq:lambda-v-compressible}), and Fig.~\ref{1d}B shows that the dependence of $v$ on $\beta$ is in good agreement with the numerically and theoretically predicted $g(\beta)$, although the theoretical form $g(\beta)=(1-\beta)^{3/4}$ is only accurate for $\beta$ close to 1. Fig. 2 shows that the square-root dependence on $E$ and $\phi$ also holds in the 2D case, but the function $g(\beta)$ is again different, and does not go to zero for $\beta>1$, due to the branching. In the Supplemental Material we perform a more rigorous derivation of Eq.~(\ref{eq:lambda-v-compressible}), showing that it is valid when the dimensionless parameter $G=E/(\mu D \rho_0)\gg1$ and $\beta$ is close to 1. We also show that mechanics-dominated growth $G\gg 1$ is relevant for any experimentally feasible parameters.

\begin{figure}
\includegraphics[width=0.45\columnwidth]{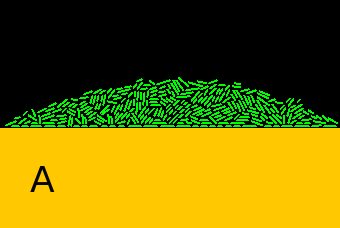}
\includegraphics[width=0.45\columnwidth]{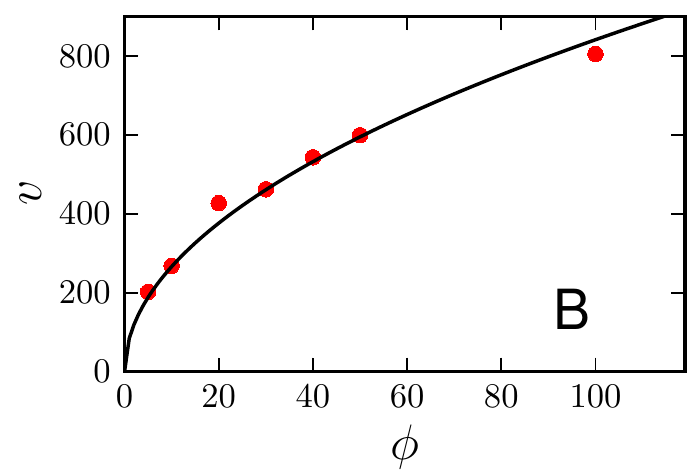}
\includegraphics[width=0.7\columnwidth]{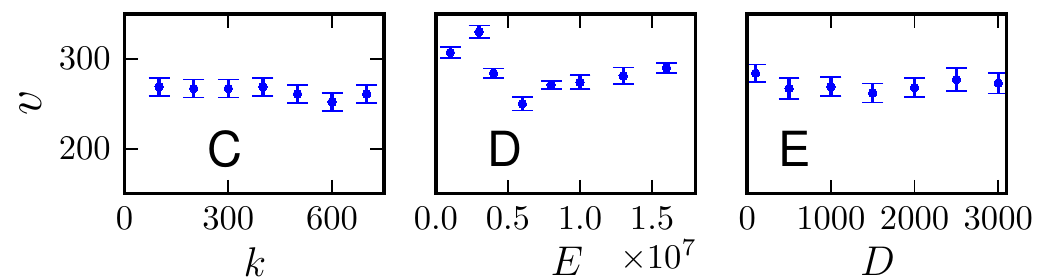}
\caption{\label{fig:3d}3D colony growth. A: snapshot. B: Speed of radial colony growth against $\phi$, with fit to $A\sqrt{\phi}$. C-E: speed against $k$, $E$ and $D$ (for $\phi=10$), showing little dependence on any of these parameters. }\label{shot3d}
\end{figure}

%effect of escape into 3rd dimension
So far, our findings are relevant to bacteria growing in monolayers. On agar plates, however, cells are observed to build up vertically in the colony centre~\cite{pipe_2008,brenner}. To probe how this additional degree of freedom affects our results, we simulate a colony growing in a vertical 2d plane $xz$  (where the $z$ axis is perpendicular to the substrate) instead of the $xy$ plane from previous simulation. We also incorporate attractive cell-cell and cell-substrate interactions, and we solve for the evolution of the nutrient field in the $z<0$ half-plane only, which models the agar gel on which growth occurs. As is apparent from the snapshot of the growth process in Fig.~\ref{fig:3d}A, cells do now escape out of the plane they start from, due to the force exerted by neighbours. The size of the colony once again grows linearly in time. However, it is not compressibility but the possibility of escape into the vertical direction which leads to linear growth.

In fact, if the bulk pressure $p_*(\epsilon)$, which builds up in a strictly two-dimensional setting, is larger than some critical pressure $p_c$, cells will escape into the $z$ dimension. As a consequence the pressure profile will saturate at $p_c$ in the bulk of the population. In our scaling argument for the speed of the front growth, we then have $v\approx [(\phi f(c_0) p_c)/(\mu \rho_0)]^{1/2}$. Figures~\ref{fig:3d}B-E show that, in contrast to the 2D case, the expansion speed $v\sim\sqrt{\phi}$ and it is independent of the consumption rate $k$, elastic modulus $E$ and the diffusion constant $D$. Note that while the radial growth is independent of $k$, the vertical growth will be affected by it.  

In conclusion, we have reported here a joint computational and analytical study of the growth of bacterial colonies where non-motile microorganisms replicate and push each other away as they grow. We find a transition between two different growth regimes, controlled by the balance between growth and uptake of nutrients. Our model differs in that from biofilm simulations \cite{kreft_bacsim_1998, idynomics} which do not explicitly model mechanical forces in the colony. We also find that the functional form of the density profile close to the bacterial edge qualitatively differs from those predicted by Fisher-Kolmogorov models, and predict that the speed at which the front propagates depends only weakly on the nutrient diffusion rate $D$, for a wide range of $D$. It would be interesting to study how  the accumulation of metabolic inhibitors \cite{hochberg_mechanism_1972}, oxygen depletion \cite{peters_oxygen_1987}, or dependence of growth rate on the distance from the agar \cite{wentland_spatial_1996} would affect our results.

{\it Acknowledgments.} We thank R. J. Allen and M. R. Evans for helpful comments on this manuscript. O.H. thanks the Deutsche Forschungsgemeinschaft (DFG) for financial support (grant A15, SFB 937). B.W. acknowledges the support of a Leverhulme Trust Early Career Fellowship.

\end{document}